\documentstyle[preprint,revtex]{aps}
\begin{document}
\draft
\preprint{ETH-L and RU96-14-B}
\preprint{October 1996}
\begin{title}
Integrabilities of the long-range t-J models\\
with twisted boundary conditions
\end{title}
\author{James T. Liu$^1$ and D. F. Wang$^2$}
\begin{instit}
$^1$ Department of Physics,
The Rockefeller University\\
1230 York Avenue, New York, NY 10021-6399, USA\\
$^2$ Institut de Physique Th\'eorique,
Ecole Polytechnique F\'ed\'erale de Lausanne\\
PHB-Ecublens, CH-1015 Lausanne, Switzerland\\
\end{instit}
\begin{abstract}
The integrability of the one-dimensional long range supersymmetric t-J
model has previously been established for both open systems and those
closed by periodic boundary conditions through explicit construction of
its integrals of motion.  Recently the system has been extended to include
the effect of magnetic flux, which gives rise to a closed chain with
twisted boundary conditions.  While the t-J model with twisted boundary
conditions has been solved for the ground state and full energy spectrum,
proof of its integrability has so far been lacking.  In this letter we
extend the proof of integrability of the long range supersymmetric t-J
model and its $SU(m|n)$ generalization to include the case of twisted
boundary conditions.
\end{abstract}
\narrowtext


Solvable models have attracted attention from both the high energy
and condensed matter communities.  These models provide important examples
where it is possible to deal with many degrees of freedom without having
to resort to perturbation theory.  Interesting models in condensed matter
physics that have been solved include the short range spin model\cite{bethe},
delta-function bose gas\cite{lieb}, delta-function electron
gas\cite{flicker,yang1}, the Hubbard model\cite{wu}, the Luttinger
model\cite{mattis1}, the magnetic impurity model and the Anderson
impurity model\cite{andrei,wiegmann}.

In one dimension, ever since Haldane and Shastry independently introduced
the $1/r^2$ spin model\cite{haldane,shastry}, there has been considerable
interest in the model and its generalizations, such as the long range
supersymmetric t-J model\cite{kura,kawa,wang3,hahaldane}.
All of these models are characterized by having a ground state wavefunction
which takes on a Jastrow product form, and by having quasi-particle scattering
matrices of a very simple form, as in the continuous Calogero-Sutherland
systems describing non-relativistic quantum particles\cite{sutherland}.
In particular, the Haldane-Shastry spin model can also be identified as
a free system composed of identical particles obeying Haldane's
generalized Pauli principle\cite{haldane2}, and obeying a generalized
statistical distribution function at finite temperature\cite{yongshi}.
In 1992, Gebhard and Ruckenstein introduced the long range Hubbard model,
in which the electrons are described by the $1/r$ Hubbard model.  It is
noteworthy that this $1/r$ Hubbard model is integrable for any on-site
energy; the full energy spectrum and thermodynamics have been solved
explicitly\cite{gebhard}.  At half-filling and in the limit of large
interaction, this model reduces to the $SU(2)$ Haldane-Shastry spin chain.
For less than half-filling, but still in the limit of $U=\infty$, the
system remains characterized by eigenfunctions of a Jastrow product form.

Recently there has been considerable interest in adding magnetic flux to
the Haldane-Shastry type models.  For a one-dimensional ring threaded by
flux, this reduces to the problem of incorporating twisted boundary
conditions.  A twisted version of the long-range integrable Haldane-Shastry
spin chain has been introduced, and was solved in the rational flux
case\cite{fukui}.  Subsequently, this was generalized to the case of the
long range t-J model with twisted boundary conditions \cite{fukui2,liuwang}.
In particular, it was shown that the irrational flux case can be treated
identically \cite{liuwang}, indicating that there is no essential
difference between rational and irrational flux.  Based on the exact
solutions, it is natural to expect that the long range models remain
integrable despite the twisted boundary conditions.  However until now
this has remained an open problem.  In this letter, we provide a proof of
the integrability of the long range t-J model and its $SU(m|n)$
generalization with twisted boundary conditions by explicitly constructing
an infinite number of simultaneous constants of motion.  This construction
is a straightforward extension of the methods used in the absence of flux
\cite{poly1,fowler,brink,itz,wang1,wang2}, and is motivated by the mapping
of the closed ring onto an equivalent open system where the flux is
manifested in twisted boundary conditions.  A further consequence of
this mapping is that it yields a unified treatment of the integrability
of both the open and closed chains.

Because of the subtleties involved with introducing magnetic flux into a
model with long range interactions, we follow the procedure of \cite{fukui},
and start with an open chain which is subsequently closed through
appropriate boundary conditions to form a ring of $N$ sites.  The
Hamiltonian of the $SU(1|2)$ t-J model on this open lattice takes the form
\begin{equation}
H_0={1\over2} P_G \sum_{\alpha\ne\beta}
{1\over (q_\alpha-q_\beta)^2} 
\times [-\sum_\sigma (c_{\alpha \sigma}^\dagger
c_{\beta \sigma}^{\vphantom{\dagger}} +h.c.)
+ \hat P_{\alpha,\beta} - (1-n_\alpha)(1-n_\beta) ] P_G,
\end{equation}
where $P_G$ is the Gutzwiller projection onto single occupancy and the
operators $c$ and $c^\dagger$ are the usual fermion operators that
satisfy standard fermionic anti-commutation relations.  The summation
over $\sigma$ is to sum over all the spin components of the electrons,
$\sigma=\uparrow,\downarrow$ for $SU(1|2)$.  The spin exchange operator
$\hat P_{\alpha,\beta}$ is given by
\begin{equation}
\hat P_{\alpha,\beta} = \sum_{\sigma^{\vphantom{\prime}}} \sum_{\sigma'}
c_{\alpha\sigma}^\dagger
c_{\alpha\sigma'}^{\vphantom{\dagger}} c_{\beta\sigma'}^\dagger
c_{\beta\sigma}^{\vphantom{\dagger}},
\end{equation}
and the electron density operator is
$n_\alpha=\sum_\sigma c_{\alpha\sigma}^\dagger
c_{\alpha\sigma}^{\vphantom{\dagger}}$.
The lattice permutation form of this Hamiltonian may be made explicit by
introducing the {\it graded} permutation operator
$\Pi_{\alpha,\beta}^{\nu,\nu'}$, which exchanges particles of species
$\nu$ and $\nu'$ at locations $\alpha$ and $\beta$ (where
$\nu,\nu'=0,\uparrow,\downarrow$ with 0 denoting a hole).  Written in terms
of $\Pi$, the Hamiltonian becomes
\begin{equation}
H_0=-{1\over2} P_G \sum_{\alpha\ne\beta}\sum_{\nu,\nu'}
{\Pi_{\alpha,\beta}^{\nu,\nu'}\over(q_\alpha-q_\beta)^2}P_G.
\label{eq:hamiltonian}
\end{equation}
In this form, the $SU(m|n)$ generalization is immediately obvious.

For a finite open chain with $L$ sites, integrability is achieved when the
lattice positions $q_\alpha$ lie at the roots of the $L^{\rm th}$ Hermite
polynomial \cite{poly1}.  In the limit $L\to\infty$ these roots become
equally spaced, and translational invariance is restored.  It is precisely
in this limit that it is possible to close the chain by demanding twisted
boundary conditions.  We allow a separate twist $\phi_\nu$ for each
independent species $\nu$, so that the twisted boundary conditions for a
ring of $N$ sites may be encoded by the requirement that
\begin{equation}
\Pi_{\alpha,\beta+l N}^{\nu,\nu'}=z^{lN(
\phi_{\nu^{\vphantom{\prime}}}-\phi_{\nu'})}
\Pi_{\alpha,\beta}^{\nu,\nu'},
\end{equation}
where $z$ is the primitive $N^{\rm th}$ root of unity.  Since the resulting
closed system is translationally invariant, we single out one period and
reexpress the Hamiltonian, after gauge
transformation, as $H_0 = E_0 + H$ where \cite{fukui} 
\begin{eqnarray}
H &=& - {1\over2} P_G\sum_{1\le i\ne j\le N}\sum_{\nu,\nu'}
\sum_{l=-\infty}^\infty {z^{(i-j-lN)
(\phi_{\nu^{\vphantom{\prime}}}-\phi_{\nu'})}\Pi_{i,j}^{\nu,\nu'}
\over(i-j-lN)^2}P_G\nonumber\\
&=& -{1\over2} P_G\sum_{1\le i\ne j\le N}\sum_{\nu,\nu'}
J_{\phi_{\nu^{\vphantom{\prime}}}-\phi_{\nu'}}(i-j)\Pi_{i,j}^{\nu,\nu'}P_G
\label{eq:twistedH}
\end{eqnarray}
now defines the long-range supersymmetric t-J model on a periodic ring.
The offset $E_0=\pi^2/3N$ accounts for exchanges $lN$ units apart
(which is present in $H_0$ but not in the periodic $H$), and may be
interpreted as a shift in the ground state energy from finite size
effects.
The sum over $l$ ensures the appropriate periodicity of the ring under
translations, and yields the inverse trigonometric potential \cite{liuwang}
\begin{eqnarray}
J_\phi(n) &=& \sum_l {z^{(n+lN)\phi}\over(n+lN)^2}\nonumber\\
&=&\left({\pi\over N}\right)^2{z^{\lfloor\phi\rfloor n}
[1+(\phi-\lfloor\phi\rfloor)(z^n-1)]\over\sin^2(n\pi/N)}.
\end{eqnarray}
This expression is piecewise linear and continuous in $\phi$, leading
to many remarkable
features of this model \cite{fukui,fukui2,liuwang}.  In the case of
periodic boundary conditions ($\phi=0$), the physical properties of the
supersymmetric t-J model on a uniform closed chain have been studied
previously\cite{kura,kawa,wang3,hahaldane}.

To provide a proof for the long range t-J model with twisted boundary
conditions, we are motivated by the previous results on the integrabilities
of the uniform long range t-J model with periodic boundary conditions, and
the non-uniform long range t-J model with open boundary conditions
\cite{wang1,wang2}, which were generalizations of the spin chain results
\cite{poly1,fowler,brink,itz}.  Proof of the integrability proceeds by first
mapping the species-exchange Hamiltonian, Eqn.~(\ref{eq:twistedH}), to a
lattice permutation equivalent using slave-boson techniques.  The resulting
Hamiltonian then acts on wavefunctions $\psi$ written in the form
$\psi(q_1\nu_1,q_2\nu_2,\ldots,q_N\nu_N)$, where $q_i$ and $\nu_i$
label the position and $SU(m|n)$ ``spin'' of particle $i$.  Acting on
such wavefunctions, and using the fact that $\{ q_i\}$ span the lattice
due to single occupancy, the Hamiltonian becomes
\begin{equation}
H=-{1\over2} \sum_{1\le i\ne j\le N} J_{\phi_{\nu_i}-\phi_{\nu_j}}(q_i-q_j)
M_{ij},
\end{equation}
where the particle exchange operator $M_{ij}$ is defined by
\begin{equation}
M_{ij}\psi(\ldots,q_i\sigma_i,\ldots,q_j\sigma_j,\ldots)
\equiv\psi(\ldots,q_j\sigma_i,\ldots,q_i\sigma_j,\ldots).
\end{equation}
Note that the fermionic and bosonic nature of the individual species
are fully encoded in the wavefunctions, $\psi\to \pm \psi$ under
simultaneous interchange of position and spin.  This independence of
the exchange operator from the particle statistics ensures that the
proof of integrability holds for {\it all} $SU(m|n)$ extended t-J models,
and not just for the $SU(1|2)$ case.

Based on the integrability proof for the open chain and for the ring
closed by periodic boundary conditions studied in Ref.~\cite{poly1,fowler},
we introduce the generalized operators
\begin{equation}
\pi_i =i \sum_{j\ne i} u_{\phi_{\nu_i}-\phi_{\nu_j}}(q_i-q_j) M_{ij},
\label{eq:pi}
\end{equation}
where $u_\phi(n)$ is the (twisted) periodic version of $1/r$
\begin{equation}
u_\phi(n) = \sum_l {z^{(n+lN)\phi}\over n+lN}.
\label{eq:rsum}
\end{equation}
As in the case for $J_\phi(n)$, this sum may be performed, yielding
\begin{equation}
u_\phi(n)={2\pi i\over N}{z^{\lfloor\phi\rfloor n}\over1-z^{-n}}
\label{eq:uphi}
\end{equation}
(for nonintegral $\phi$).  Note that $\phi$ enters discontinuously, with
a jump in $u_\phi(n)$ at integral values of $\phi$.  A careful treatment of
convergence issues for integral $\phi$ indicates that the actual value of
the infinite sum in Eqn.~(\ref{eq:rsum}) is the average of the values of
$u_\phi(n)$ before and after the discontinuity.  Nevertheless, for a
consistent treatment of the invariants, we take Eqn.~(\ref{eq:uphi}) as the
definition of $u_\phi(n)$ for {\it all} values of $\phi$.  A consequence of
this asymmetry is to pick a preferred ordering, thus breaking the parity
symmetry $u_{-\phi}(-n) = -u_\phi(n)$, which otherwise holds for nonintegral
$\phi$.  Nevertheless, this particular choice of ordering gives
\begin{equation}
u_0(n)={2\pi i\over N}{1\over1-z^{-n}},
\end{equation}
in agreement with previous results in the absence of flux
\cite{poly1,fowler}.  Overall, this subtle treatment of integral twists
indicates that, surprisingly enough, it is actually the {\it zero} flux
case that is exceptional; this case corresponds to working on top of the
locations of the cusps in the spectral flow itself.

Explicit evaluation of the commutators brings out a distinction between
integral and non-integral twists:
\begin{equation}
[\pi_i,\pi_j]=\cases{
{2\pi\over N}z^{\phi_{ij}q_{ij}}M_{ij}(\pi_i-\pi_j)&for $\phi_{ij}\in Z$\cr
0&otherwise,}
\end{equation}
where $\phi_{ij}=\phi_{\nu_i}-\phi_{\nu_j}$ and $q_{ij}=q_i-q_j$.  Using the
relation $(z^{\phi_{ij}q_{ij}}M_{ij})\pi_j=\pi_i(z^{\phi_{ij}q_{ij}}M_{ij})$,
valid whenever $\phi_{ij}\in Z$, the above commutator may be reexpressed in
a form similar to that of Ref.~\cite{fowler}
\begin{equation}
[\pi_i,\pi_j]=\cases{
{2\pi\over N}[z^{\phi_{ij}q_{ij}}M_{ij},\pi_i]&for $\phi_{ij}\in Z$\cr
0&otherwise.}
\end{equation}
{}From here it is obvious that the commutation results of Folwer and
Minahan \cite{fowler} are easily generalized to the present case.
Therefore the infinite set of hermitian operators
\begin{equation}
I_M= \sum_j \pi_i^M
\end{equation}
(where $M=0,1,2,\ldots$), provides a set of mutually commutating operators,
$[I_M,I_N]=0$, regardless of the individual species twists.  Note that the
commutation is trivial for nonintegral relative twists, and is basically a
consequence of the simple open chain result,
$[\pi_{0\,\alpha},\pi_{0\,\beta}]=0$, where
\begin{equation}
\pi_{0\,\alpha}=i\sum_{\beta\ne\alpha}{1\over q_\alpha-q_\beta}M_{\alpha\beta}
\end{equation}
is the corresponding open chain operator [compare with Eqn.~(\ref{eq:pi})].

For a finite open chain, it is known that the $\pi_0$ operators do not
commute with the Hamiltonian \cite{poly1},
\begin{equation}
[H_0, \pi_{0\,\alpha}] = 2i  \sum_{\beta\ne\alpha}
{1\over (q_\beta-q_\alpha)^3}.
\label{eq:comm}
\end{equation}
However, in making the system periodic, the odd exponent in
Eqn.~(\ref{eq:comm}) allows the cancellation of terms exchanging to the
left and right.  As a result, we find
\begin{equation}
[H,\pi_i]=0
\end{equation}
for the periodic case, regardless of the relative twist angles.
It is now clear that the mutually commuting set of operators, $\{I_M\}$
for $M=0,1,2,\ldots$, provide explicit constants of motion of the
Hamiltonian $H$, and hence proves the integrability of the long range
$SU(m|n)$ t-J model on a ring (for either twisted or untwisted boundary
conditions).

In conclusion, we have provided a proof for the integrability of the
long range t-J models with twisted boundary conditions by explicitly
constructing an infinite set of mutually commuting constants of motion.
This proof generalizes previous results for rings without flux, and makes
use of the viewpoint that the closed chain is simply a periodic version
of the open system.  A consequence of this similar treatment for both
closed and open chains is the demonstration that the key property behind the
integrability of these models is simply the permutation nature of the system.
These results have filled a gap in that the integrability condition for the
twisted t-J model was as yet unknown, in spite of the fact that several
thorough studies of the long range model in the presence of flux have
been provided.

\bigskip

We wish to thank C. Gruber, H. Kunz, R. Khuri and H. C. Ren
for stimulating discussions.  In addition, one of us (DFW)
wishes to thank P. Coleman for encouragement.
This work was supported in part by the U.~S.~Department of Energy under
grant no.~DOE-91ER40651-TASKB, and by the Swiss National Science Foundation.

\end{document}